\PassOptionsToPackage{table}{xcolor}
\documentclass[sigconf, nonacm]{acmart}

\usepackage{pvldb}% VLDB/PVLDB formatting overrides, see pvldb.sty for details.
\renewcommand\vldbdoi{XX.XX/XXX.XX}
\renewcommand\vldbpages{XXX-XXX}
\renewcommand\vldbavailabilityurl{https://github.com/delftdata/agenti-ai-guarantees} 

\usepackage{booktabs}   % Required for \specialrule and table rules
\usepackage{multirow}   % Required for merged cells
\usepackage{tabularx}   % Required for wrapping table columns
\usepackage{pifont}     % Required for \cmark/\xmark table glyphs
\newcommand{\cmark}{\ding{51}}
\newcommand{\xmark}{\ding{55}}
\usepackage{todonotes}  % Required for \todo commands
\usepackage{tikz}
\usetikzlibrary{positioning, arrows.meta, calc}
\usepackage{graphicx}
\usepackage{comment}
\usepackage{makecell}
\usepackage{cleveref}
\usepackage{xspace}
\usepackage{enumitem}

\AtBeginDocument{%
  }

\newcommand{\para}[1]{\vspace{1mm}\noindent\textbf{#1}}

\newcommand{\asterios}[1]{\todo[inline, color=red!20]{Asterios: #1}}

\newcommand{\subt}[1]{{\small\itshape #1}} % or whatever styling you intend for subtitles

\begin{document}

\title[{Tracking State Footprints: How Agents Can Transact}]{Tracking State Footprints: How Agents Can Transact}

% \subtitle{[Vision]}

%%
%% The "author" command and its associated commands are used to define
%% the authors and their affiliations.

%% Grouped author entry to save space
\author{Oto Mraz \, Rares Şerban \,  Kyriakos Psarakis\texorpdfstring{$^\dagger$}{} \, Burcu Kulahcioglu Ozkan \, Asterios Katsifodimos}
\affiliation{%
  \institution{$^\dagger$Ververica GmbH \qquad Delft University of Technology}
  % \city{Delft}
  % \country{Netherlands}
}
% \email{\{o.m.mraz, r.c.serban, b.ozkan, a.katsifodimos\}@tudelft.nl, kyriakos.psarakis@ververica.com}

%% Previous separate author setup (kept for reference):
% \author{Oto Mraz}
% \orcid{1234-5678-9012}
% \affiliation{%
%   \institution{Delft University of Technology}
%   \city{Delft}
%   \country{Netherlands}}
% \email{o.m.mraz@tudelft.nl}
% 
% \author{Kyriakos Psarakis}
% \affiliation{%
%   \institution{Ververica GmbH}
%   \city{Delft}
%   \country{Netherlands}}
% \email{kyriakos.psarakis@ververica.com}
% 
% \author{Rares Şerban}
% \affiliation{%
%   \institution{Delft University of Technology}
%   \city{Delft}
%   \country{Netherlands}}
% \email{r.c.serban@student.tudelft.nl}
% 
% \author{Burcu Kulahcioglu Ozkan}
% \affiliation{%
%   \institution{Delft University of Technology}
%   \city{Delft}
%   \country{Netherlands}}
% \email{b.ozkan@tudelft.nl}
% 
% \author{Asterios Katsifodimos}
% \affiliation{%
%   \institution{Delft University of Technology}
%   \city{Delft}
%   \country{Netherlands}}
% \email{a.katsifodimos@tudelft.nl}

\renewcommand{\shortauthors}{Mraz et al.}
\renewcommand{\vldbauthors}{Oto Mraz, Rares Şerban, Kyriakos Psarakis, Burcu Kulahcioglu Ozkan, and Asterios Katsifodimos}

%%
%% The abstract is a short summary of the work to be presented in the
%% article.
\begin{abstract}
Can AI agents transact? We argue that they must: as multi-agent systems (MASs) increasingly write code, deploy infrastructure, modify databases, and call web services, lost updates or stale reads can be catastrophic. Although MASs increasingly execute plans in parallel, current orchestrators do not track the state that agents read and write. As a result, concurrency anomalies manifest even in simple coding tasks. We frame MAS coordination as a data management problem and propose to describe agents by their \emph{state footprint}: the state they read and write across their own local context and state, as well as the state of the orchestrator and external systems. We posit that MASs require guarantees similar to those of databases, but providing them raises new challenges and opportunities: unlike database transactions, agents do not read from a fixed schema or an isolated snapshot, and cannot be replayed deterministically upon failure. They can, however, resolve conflicts semantically instead of aborting, enabling new forms of concurrency control and conflict resolution. Towards agents that can transact, we outline a vision for next-generation agent orchestrators and transactional interfaces for external systems to participate in agentic transactions.

% Multi-agent systems (MASs) employ multiple LLM agents to solve complex tasks, such as coding, data analysis, and reporting. The main observation behind this paper is that MASs share foundational challenges with classical databases, including communication, coordination, and state management. While MASs benefit from specialized, parallel execution, the inherent non-determinism of LLMs inhibits their reliability. As we show, parallelism without data-dependency tracking breaks correctness without saving tokens. In this paper, we frame MAS coordination problems as a data management problem. We demonstrate how coordination failures manifest even in simple coding tasks and propose mechanisms to address them. By systematically tracking the state footprint of MAS, next generation MAS orchestrators could open doors to favorable atomicity, consistency and isolation, durability, and fault tolerance guarantees. We further argue that the external systems agents act upon must be redesigned to support these guarantees. Finally, we highlight fertile research opportunities and open problems aiming salvaging intermediate results and minimizing end-to-end task duration and cost, while improving accuracy.
\end{abstract}

\maketitle

%%% do not modify the following VLDB block %%
%%% VLDB block start %%%
\vldbtopmatter
%%% VLDB block end %%%

\section{Introduction}

Recent advances in LLM agents enable collaborative multi-agent systems (MASs) to solve complex tasks, including bug fixing~\cite{qian2024chatdev,bouzenia2025repairagent,dang2026chatdev}, deployment \& incident management~\cite{malik2026autonomous,ko2026seven}, or supply chain control~\cite{yin2025rethinking,ko2026seven}. In a typical MAS, a planner agent decomposes a user's goal into subtasks, and an orchestrator dispatches them to specialized worker agents, often in parallel to finish faster. The workers read and modify shared state along the way: a code repository, a database, or a third-party service. Yet, MASs remain unreliable in practice~\cite{xie2024osworld,wang2024survey}. The usual explanation lies in the models: LLM agents are nondeterministic, prone to hallucinations that cascade into downstream agents~\cite{hong2024metagpt,qian2024chatdev,darwish2025mitigating,shi2025mitigating} or trigger endless thought loops~\cite{huang2024understanding}. In this paper, we argue that this unreliability lies not only in the models, but also in how their work is coordinated.

Consider a team of coding agents refactoring a codebase. A planner splits the work into subtasks: one agent changes a core component, others update the code that uses it. Every agent completes its subtask successfully, yet the merged code fails its tests: some agents updated callers against the old version of the component, and two concurrent agents edited the same file, one silently overwriting the other. A better LLM would not have prevented either failure, since each agent acted correctly in isolation. The issue is that most of today's orchestrators track \emph{control flow}, i.e., which task runs after which, but not \emph{data flow}, i.e., which \textit{state} each task reads and writes~\cite{hong2024metagpt,dang2026chatdev}. The ML community has greatly improved planning~\cite{erdogan2025plan,wang2023plan,huang2024understanding}, collaboration~\cite{zhou2026multi,yang2026agentnet}, speed, context capacity, and token costs, but these advances govern what agents decide, not how their access to state is coordinated. The data management community, in contrast, has long-established paradigms for concurrency, fault tolerance, and consistency~\cite{bailis2013highly,yang2022oceanbase,ailijiang2019dissecting}, yet these are 
not systematically applied to agentic orchestration. In this paper, we examine MAS coordination through a data management lens.

Our central thesis is that reliability must come from the orchestrator, not the models alone: plans and actions come from LLMs and may be wrong, but the guarantees must come from a deterministic component. This component should record agents' state accesses and enforce ordering regardless of how the model behaves. State is scattered across three scopes: the agent's own context and local files, the orchestrator's plan and task status, and external systems (code repositories, databases, web services). In this paper, we describe a task by its \textit{footprint}\footnote{Picture an agent walking through local and external state, inspecting or modifying it along the way; the footprint is the trail it leaves. The term follows separation logic, where a program's footprint is the set of locations it accesses~\cite{reynolds2002separation}.}: the state it reads and writes across all three scopes. We outline how footprints underpin the guarantees that databases provide: atomicity, consistency, isolation, durability.

Multi-agent orchestration presents new challenges. Unlike a database transaction, an agent does not read from a fixed schema but from whatever enters its context window. In addition, its execution cannot be replayed deterministically in case of failure. At the same time, agents bring a capability that transactions lack: instead of aborting on a conflict, they can inspect it and perform semantic conflict resolution. MASs bear similarities to databases in the \textit{guarantees} they need, but not in the \textit{mechanisms} that provide them.

\vspace{1mm}
\noindent This paper makes the following contributions:

\begin{itemize} [leftmargin=*]
    \item We show that MAS coordination failures can stem from untracked data dependencies: orchestrators order tasks, but not the state they read and write (\S\ref{sec:background}). We characterize anomalies on four coding tasks (\S\ref{sec:results}).
    \item We identify three scopes of state in MASs (agent-local, orchestrator, and external) that differ in who can observe, revert, and reproduce them in case of conflict or failure (\S\ref{sec:background:data-management}).
    \item We propose to track agentic tasks by their \emph{footprints}: the state they read and write across all scopes, capturing conflicts that span scopes and may emerge dynamically during execution (\S\ref{sec:footprints}).
    \item We show how state footprints and provenance can underpin ACID guarantees, fault tolerance, and reuse, and identify open research questions each raises (\S\ref{sec:next_gen_agentic_orchestration}).
    \item We outline how external systems can become \emph{agent-ready} by exposing transactional interfaces that let orchestrators stage, validate, and revert state effects (\S\ref{sec:agent-ready}).
\end{itemize}

\section{A Brief Introduction to MAS}\label{sec:Preliminaries}

\begin{figure}[t]
\centering
\includegraphics[width=0.9\columnwidth]{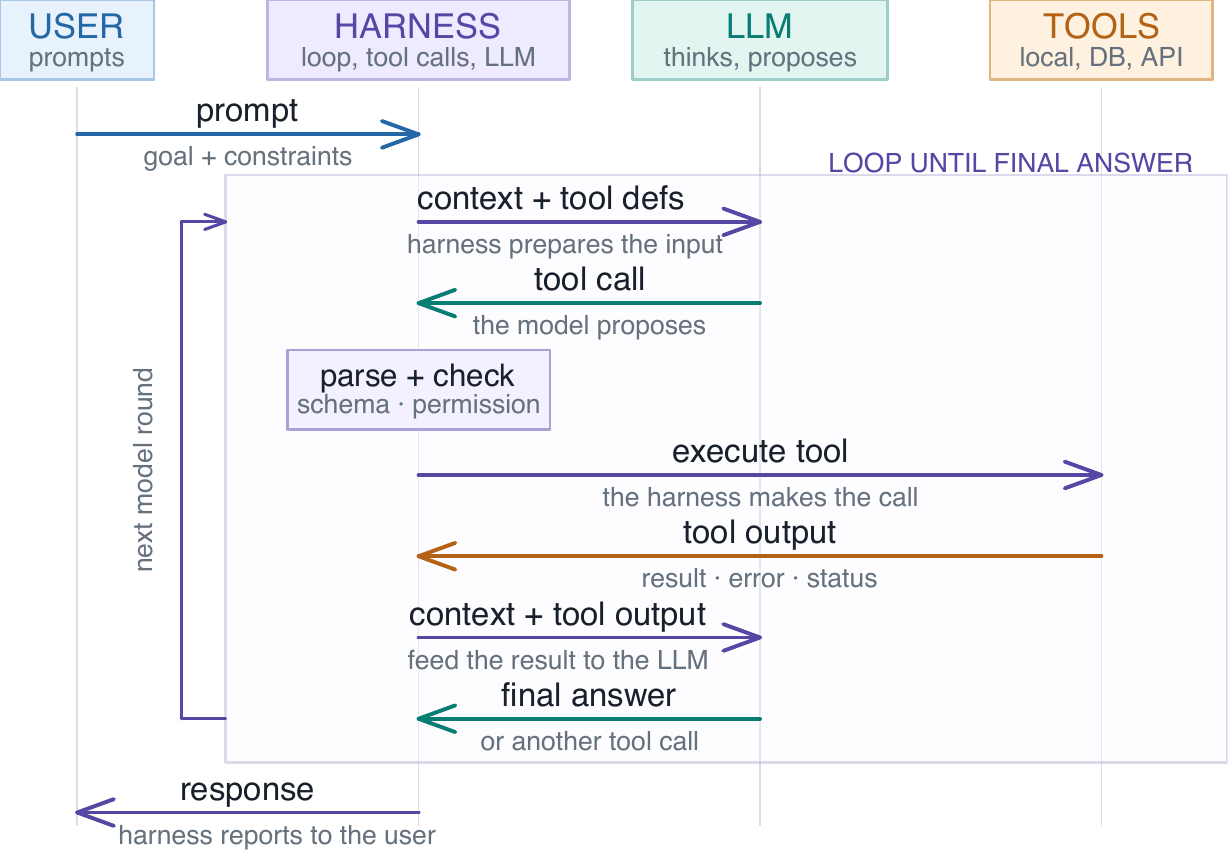}
\vspace{-2mm}
\caption{The main "loop" of a single agent. The LLM \emph{proposes} tool calls, while the harness controls the loop, manages state, validates permissions, and executes tools.}
\label{fig:agentic_architecture}
\Description{Agentic system architecture}
\vspace{-2mm}
\end{figure}

%We begin by describing the foundational components of agents and their historical evolution.

\para{Architecture.} Within a single agent (\Cref{fig:agentic_architecture}), execution proceeds via a closed loop between the LLM and runtime harness. The LLM is responsible for reasoning, decision-making, tool selection, and output evaluation. As LLMs operate strictly on text, they rely on a \emph{harness}, a wrapper around the LLM(s), that manages state, invokes tools, and executes code~\cite{cheng2026exploring,yang2024if}. The harness parses the output of an LLM, looking for special tags that LLMs are trained to produce in their output~\cite{li2026xgrammar}. The harness invokes the requested tools and routes their results back to the LLM. 

An MAS coordinates an \textit{ensemble} of such agents to solve a user-specified task. In an MAS, individual LLM agents assume dedicated roles (e.g., coding, testing, deploying)~\cite{hong2024metagpt,he2025llm,qian2024chatdev}. To orchestrate execution, a dedicated planner decomposes the high-level goal into structured subtasks with constraints~\cite{zhou2026multi,wang2023plan,fang2025comprehensive,bajaj2026position}. Plans manage a speed vs. correctness tradeoff and may be dynamically modified on the fly. The literature mainly distinguishes sequential and DAG-based topologies. A \emph{sequential} plan executes subtasks strictly one by one. While safest to coordinate, sequential plans are also the slowest~\cite{qian2024chatdev}. DAG-based topologies use scatter-gather patterns to run subtasks in parallel, merging the artifacts at the end~\cite{hong2024metagpt}. 

\para{Historical evolution.}\label{sec:orchestrator_generations}
Following AutoGPT~\cite{firat2023if}, early MAS frameworks (2023--24) such as ChatDev~\cite{qian2024chatdev} and LangChain~\cite{chase2022langchain} executed static plans with rigid coordination~\cite{fang2025comprehensive}: an intermediate failure required a complete restart~\cite{huang2024understanding}. State management and idempotency were left to the user. Second-generation runtimes (2024--26) such as LangGraph~\cite{langgraph2024}, MetaGPT~\cite{hong2024metagpt}, AutoGen~\cite{wu2024autogen}, and ChatDev 2.0~\cite{dang2026chatdev} introduced DAG-based execution with dynamic replanning~\cite{huang2024understanding}, task verification, and self-healing. They can detect and resolve many failures without human intervention~\cite{fang2025comprehensive,yang2025unlocking,yang2026agentnet,asthana2026structured}, but without guarantees about the \textit{state} their tasks share. Emerging frameworks like EvoFlow~\cite{zhang2025evoflow} and EvoAgentX~\cite{wang2025evoagentx} evolve both execution plans and agent roles at runtime. Combined with the shift towards multi-modal agents~\cite{dang2026chatdev}, this flexibility exacerbates the challenges of shared state management, coordination, recoverability, and reconciliation. Here we argue that the next generation of runtimes (2027+) \textit{must provide correctness guarantees} via explicit concurrency control and data management principles.

\section{Multi-Agent Coordination as a Data Management  Problem}\label{sec:background}
%(check todo.md)

%Running example: Code refactoring task

MAS coordination is fundamentally a data management challenge: coordination correctness depends not only on \emph{control-flow} (task ordering and temporal scheduling) but also on the \emph{data-flow}: data dependencies~\cite{DBLP:journals/toplas/ShashaS88}, output visibility and isolation~\cite{DBLP:conf/isca/HerlihyM93, DBLP:journals/computer/AdveG96, adya1999weak} across agents. Agent orchestrators so far primarily capture control flow~\cite{hong2024metagpt,dang2026chatdev,langgraph2024}, neglecting data dependencies arising from conflicting reads and writes to shared state. Agents may thus act on a stale or inconsistent view of the state, causing failures that control-flow ordering alone cannot prevent. We illustrate the coordination problems on a seemingly simple agentic coding example, which shows that coordination and isolation schemes arise well beyond databases. %are present in tasks unrelated to databases.

%We need a data management layer that complements control-flow coordination, tracks data dependencies across global and agent-local state, and mediates the shared-state accesses.

% The data management layer will provide fault tolerance through state recovery, and provide workflow execution guarantees. 

\begin{example}[API refactoring task]\label{motiv-ex}
Consider a SWE-bench-style \cite{jimenez2024swebench} refactoring task to remove the deprecated \texttt{providing\_args} argument from \texttt{Signal} API~\cite{django-ticket-31327} in Django~\cite{django}. The constructor used to accept an argument \texttt{providing\allowbreak{}\_args}, removed in v4.0. The refactoring task has a fanout structure: one edit to the core \texttt{Signal} definition (remove the constructor argument and delete the stored \texttt{providing\_args} attribute), along with the edits of \texttt{Signal} instantiations in subpackages (\texttt{db}, \texttt{contrib}, \texttt{core}), and test updates.
\end{example}

\para{Agentic workflow.} Figure~\ref{fig:task-dag} depicts the task workflow as a DAG that encodes task (control flow) dependencies. A supervisor agent decomposes the issue into a task editing the core \texttt{Signal} definition (T0), three concurrent tasks (T1--T3), and dispatches them. The runtime executes the graph over shared state (code on a local filesystem). %(LangGraph's supervisor pattern; AutoGen's GroupChat manager; CrewAI's hierarchical process; Claude Code's lead agent spawning subagents) 
Worker T0 modifies the constructor signature and deletes the stored attribute. Workers T1--T3 each own an independent caller subpackage and its tests. %, handling \texttt{db signals}, \texttt{auth callers}, and \texttt{handlers} respectively.
%\texttt{db signals}, \texttt{auth callers}, and \texttt{handlers} respectively. 
%\texttt{signals.py}, \texttt{contrib/auth}, and \texttt{handlers} respectively. 
The workers' write sets initially appear disjoint, but parallel execution triggers two subtle anomalies:

\definecolor{writerFill}{HTML}{EEEDFE} \definecolor{writerLine}{HTML}{534AB7}
\definecolor{turnFill}{HTML}{E1F5EE}   \definecolor{turnLine}{HTML}{0F6E56}
\definecolor{gateFill}{HTML}{FAEEDA}   \definecolor{gateLine}{HTML}{854F0B}
\definecolor{effFill}{HTML}{F1EFE8}    \definecolor{effLine}{HTML}{5F5E5A}
\definecolor{unplanned}{HTML}{A32D2D}

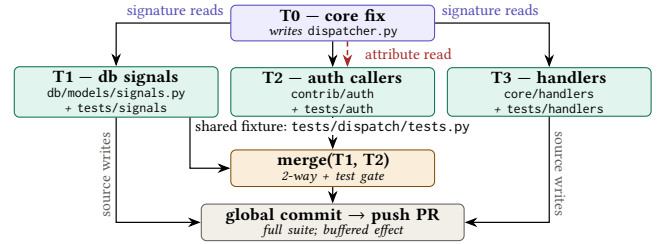
\begin{figure}[t]
\centering
\resizebox{0.48\textwidth}{!}{

\begin{tikzpicture}[
  font=\large,
  task/.style={draw, rounded corners=3pt, align=center, line width=0.6pt,
               text width=40mm, inner ysep=1.8pt, inner xsep=2pt},
  writer/.style={task, fill=writerFill, draw=writerLine},
  turn/.style  ={task, fill=turnFill,   draw=turnLine},
  gate/.style  ={task, fill=gateFill,   draw=gateLine},
  effect/.style={task, fill=effFill,    draw=effLine, text width=52mm},
  dep/.style={-{Stealth[length=2.6mm]}, line width=0.7pt},
  bad/.style={dep, densely dashed, draw=unplanned, line width=0.9pt}
]
% Rows -- tight vertical spacing and economic padding
\node[writer] (T0) at (0,0)
  {\textbf{T0 --- core fix}\\[-3.5pt]\subt{writes \texttt{dispatcher.py}}};
\node[turn] (T2) at (0,-1.4)
  {\textbf{T2 --- auth callers}\\[-3.5pt]\subt{\texttt{contrib/auth}}\\[-3.5pt]\subt{+ \texttt{tests/auth}}};
\node[turn] (T1) at (-4.4,-1.4)
  {\textbf{T1 --- db signals}\\[-3.5pt]\subt{\texttt{db/models/signals.py}}\\[-3.5pt]\subt{+ \texttt{tests/signals}}};
\node[turn] (T3) at (4.4,-1.4)
  {\textbf{T3 --- handlers}\\[-3.5pt]\subt{\texttt{core/handlers}}\\[-3.5pt]\subt{+ \texttt{tests/handlers}}};
\node[gate] (MG) at (0,-2.95)
  {\textbf{merge(T1, T2)}\\[-3.5pt]\subt{2-way + test gate}};
\node[effect] (GC) at (0,-4.05)
  {\textbf{global commit $\rightarrow$ push PR}\\[-3.5pt]\subt{full suite; buffered effect}};

% ---- Planned signature reads ----
\draw[dep] (T0.south) -- (T2.north);
\draw[dep] (T0.west) -| (T1.north);
\draw[dep] (T0.east) -| (T3.north);

% ---- Unplanned attribute read (red) ----
\draw[bad] ($(T0.south)+(0.3,0)$) -- ($(T2.north)+(0.3,0)$);

% ---- Fixture writes into merge sides ----
\draw[dep] ($(T1.south)+(1.5,0)$) |- (MG.west);

% ---- Source writes: straight down from outer-bottom, single turn into GC ----
\draw[dep] (T1.south) |- (GC.west);
\draw[dep] (T3.south) |- (GC.east);

% ---- Merge output + T2 straight down ----
\draw[dep] (MG.south) -- (GC.north);
\draw[dep] (T2.south) -- (MG.north);

% ---- Labels ----
\node[writerLine, font=\normalsize, anchor=south, inner sep=1.5pt] at (-3.2,0.02) {signature reads};
\node[writerLine, font=\normalsize, anchor=south, inner sep=1.5pt] at (3.2,0.02) {signature reads};
\node[unplanned, font=\normalsize, anchor=west] at (0.55,-0.62) {attribute read};
\node[font=\normalsize, fill=white, inner sep=1.0pt] at ($(T2.south)!0.30!(MG.north)$) {shared fixture: \texttt{tests/dispatch/tests.py}};
\node[effLine, font=\normalsize, anchor=south, rotate=90, inner sep=1.5pt] at ($(T1.south)+(-0.06,-1.15)$) {source writes};
\node[effLine, font=\normalsize, anchor=south, rotate=-90, inner sep=1.5pt] at ($(T3.south)+(0.06,-1.15)$) {source writes};

\end{tikzpicture}

}
\vspace{-6mm}
\caption{DAG for a refactoring task. The updated \texttt{Signal} definition (T0) is read concurrently by three tasks (T1–T3). Solid edges are planned reads of T0's new signature; the dashed red edge is an unplanned read of the deleted attribute. T1 and T2 write a shared test file, with the write-write conflict resolved by a merge. All outputs are combined at a global commit, which validates the result before releasing the PR.}
\label{fig:task-dag}
\Description{DAG for the \texttt{providing\_args} removal task.}
\vspace{-5mm}
\end{figure}

\begin{figure*}[t]
\centering
\includegraphics[width=\textwidth]{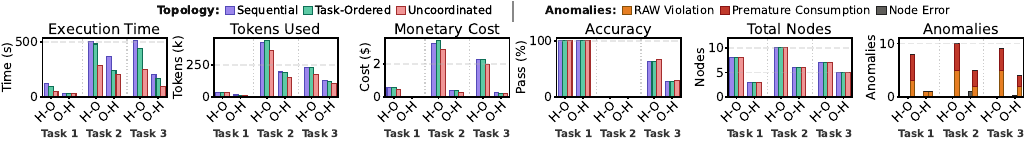}
\vspace{-5mm}
\caption{Performance of harness-authored plans for Tasks 1-4. The ticks denote the planner-agent pairs (Haiku (H), Opus (O)).}
\label{fig:system_evaluation_metrics}
\vspace{-2mm}
\end{figure*}

\para{A write-write conflict, silently reconciled.} While T1 and T2's assigned subpackages are disjoint, both write to a shared fixture, \texttt{tests/dispatch/tests.py}, which the merge node reconciles. The task whose fixture is shadowed never executes its intended setup. The tests still pass, having silently run against the wrong fixture or missing intended test coverage. In contrast, T3 has no overlap and proceeds directly to the global commit node. The global commit integrates all subpackages, the merged test file, and validates the entire repository. If successful, the workflow's only external side effect is released: pushing the branch and opening the PR.
%Note that most of the changes that arrive here need no reconciliation: the disjoint writes (T3's files, each slice's own source, and merged tests) compose by simple union, guaranteed conflict-free by the footprint declarations before anything runs. The agent runs the tests, and only if they pass does it release the single external side effect: pushing the branch and opening the PR. 

% TODO: This can be used both to motivate data-dependency declaration, also fault tolerance and dynamic plans
\para{An unplanned dependency, invisible until failure.} The dashed red edge marks an unplanned dependency of T2 on T0. Besides modifying the constructor signature, T0 also deletes the \texttt{providing\_args} attribute from \texttt{Signal}, which the plan did not expect any agent to depend on. Yet, T2 reads this attribute in a debug path, triggering a dynamic dependency on state removed by T0. Lacking a means to track this dependency, the conflict goes unnoticed until the final test suite fails, despite each task completing successfully.

%Handling this correctly requires explicit coordination across turns: tracking the dependencies between their reads and writes, and recovering the work invalidated when a dependency changes.   

% \subsection{Control-Flow Coordination}
% \label{sec:background:control-flow}
%The most interesting of them all. Scatter-gather, shared state, "stages" for data parallelism, "Thread.joins" or "async/await" for task parallelism.

% Control-flow coordination governs how work is decomposed and scheduled in a workflow. MAS orchestrators at the moment coordinate tasks via an execution topology defining task ordering as discussed in \Cref{sec:Preliminaries}. However, they do not track shared state conflicts. State in this case is files in a common filesystem. This can lead to anomalous or erroneous workflow executions as seen in the workflow in Example~\ref{motiv-ex}: T0 forks into concurrent fan-out tasks that join at a barrier (merge, global commit). 

\para{Write-After-Write (WAW) violation.} T1 and T2 collide on the shared test fixture \texttt{tests/dispatch/tests.py}. As the plan assumes their write sets are disjoint, it does not detect the overlapping targets prior to dispatch, causing one to overwrite the other silently. % TODO silent problem with both tests added

\para{Read-After-Write (RAW) violation.} T2 correctly executes after T0 terminates, inheriting the updated constructor signature. However, control-flow ordering only guarantees that T0 has finished, not that its side effects are semantically compatible with downstream tasks. T0's removal of the stored \texttt{providing\_args} is one such side effect. T2's debug path expects this field, but because this data dependency is unrepresented in the graph, the conflict bypasses ordering checks and surfaces only when post-merge tests execute the missing path. A RAW violation is similar to a \emph{stale read} in classical concurrency control, except that this data dependency is never expressed in a way for the workflow to catch.

While WAW and RAW violations resemble those in traditional databases, where dependency problems are typically resolved by ordering, \textit{the violations in MAS require mechanisms beyond task ordering}.
In the RAW example, T2 ran after T0, yet the failure occurred regardless. The cause was not \emph{when} T2 ran, but whether T0's write was checked against a dependency. Similarly, fixing WAW requires not simply ordering tasks but dependency tracking and a merge strategy to reconcile two concurrent writes. % into a single intended result.

\section{The Cost of Ignoring Dependencies}
%\section{Agents in Action}
\label{sec:results}

We empirically evaluate how the MAS coordination issues in \S\ref{sec:background} manifest on four agentic coding tasks originating from SWE-Bench \cite{jimenez2024swebench} using two LLMs and four execution topologies. To account for LLM nondeterminism, we report the averages over five runs. %The key questions we want to answer are:
%\begin{itemize} [leftmargin=*]
%    \item Do isolation anomalies arise in parallel workflows in practice?
%    \item What are the trade-offs between accuracy, execution speed, and token costs of different topologies?
%    \item How can a failed plan be repaired for a subsequent attempt?
%\end{itemize}

\para{Models.} We evaluate Claude Haiku 4.5 and Claude Opus 4.8~\cite{anthropic2024claude3} alternating between planner and worker roles. To isolate baseline coordination behavior, we disable dynamic replanning. We expect the anomalies to carry over to dynamic plans, since each replanning step yields a new static plan that executes in the same way.

\para{Tasks.} We employ four coding tasks provided as GitHub issues~\cite{jimenez2024swebench}. The agents create a fix, checked by a set of tests. \emph{Task 1} fixes a pattern matching bug with localized symptoms, root-cause diagnosis, and a list of files to fix. It enables a static plan a priori. \emph{Task 2} resolves a mathematical error, only based on symptoms. Without systematic code analysis, the orchestrator must guess where the problem actually lies. \emph{Task 3} repairs a media object merge fault. Symptoms and diagnosis are provided, but dynamic inter-agent communication is needed for a successful fix. \emph{Task 4} is the Django API refactoring task from \Cref{fig:task-dag}. We supply a human-authored plan to test dependency management during execution and recovery.

%We deliberately select software engineering tasks as the most well-understood agentic use case. The anomalies we study, however, arise wherever agents share mutable state (\Cref{sec:looking_ahead}).
%burcu{How about we flatten the itemized list?}
%\begin{itemize} [leftmargin=*]
%\item \textbf{Task 1 (Easy):} Regex pattern matching with localized symptoms, root-cause diagnosis, and a list of files to fix. Enables a static plan a priori. % , correct 
%\item \textbf{Task 2 (Hard):} Mathematical bug fixing, only providing symptoms. Without systematic code analysis, the orchestrator must guess where the problem actually lies.
%\item \textbf{Task 3 (Medium):} Media object merge fault where symptoms and diagnosis are provided, but requires dynamic inter-agent communication for a successful fix.
%\item \textbf{Task 4 (Medium):} The Django API refactoring task from \Cref{fig:task-dag}. We supply a human-authored plan to test dependency management during execution and recovery.
%\end{itemize}

\para{Topologies.} Tasks execute under four predefined execution topologies: \emph{Sequential} (strict serialized execution), \emph{Uncoordinated} (aggressive concurrency that relaxes dependencies to minimize execution time), \emph{Task-Ordered} (concurrent dispatch while enforcing task precedence), and \emph{Conflict-Ordered} (concurrent execution that additionally enforces data dependencies and conflict isolation). \Cref{tab:topology-properties} summarizes the coordination guarantees each topology provides.

\begin{table}[t]
\centering
\rowcolors{2}{white}{gray!20}
\caption{Coordination guarantees of the execution topologies.}
\vspace{-2mm}
\label{tab:topology-properties}
\resizebox{\columnwidth}{!}{%
\begin{tabular}{l c c c}
\specialrule{.1em}{.05em}{.05em}
\textbf{Execution Topology} & \textbf{Task-Ordered} & \textbf{Parallel} & \textbf{Conflict-Ordered} \\
\specialrule{.1em}{.05em}{.05em}
Sequential       & \cmark & \xmark & \xmark \\
Uncoordinated    & \xmark & \cmark & \xmark \\
Task-Ordered     & \cmark & \cmark & \xmark \\
\rowcolor{green!10}
Conflict-Ordered & \cmark & \cmark & \cmark \\
\specialrule{.1em}{.05em}{.05em}
\end{tabular}%
}
\vspace{-4mm}
\end{table}

\para{Metrics.} We measure efficiency via \emph{execution time}, total \emph{tokens used}, and monetary \emph{cost}. The cost comprises input \& output tokens, and cache reads \& writes priced \$0.1--5 and \$0.5--25 per 1M tokens for Haiku and Opus~\cite{anthropic_claude_pricing}. Correctness is assessed by \emph{accuracy} and the anomalies: $i)$~\emph{node errors} (an agent fails on its own input), $ii)$~\emph{premature consumptions} (reading partial outputs prior to task completion), $iii)$~\emph{RAW} violations, and $iv)$~\emph{WAW} violations as described in \S\ref{sec:background}.

%\subsection{The Cost of Aggressive Parallelism}
%\vspace{2mm}
%\noindent We now turn to the experiment results.

\para{Missing data dependencies.} As the harness increases parallelism, RAW violations and premature consumptions escalate due to the unmodeled data dependencies (\Cref{fig:system_evaluation_metrics}). Opus constructs more robust plans than Haiku, yielding fewer nodes and thus fewer anomalies. To isolate planner quality from worker agent strength, we also tested a Haiku-Haiku setup (not plotted for lack of space). While reducing time and token costs, its anomaly rates closely mirror Haiku-Opus, confirming the need for sound data dependency management. Task complexity dictates resilience to execution anomalies. In the easiest Task~1, downstream agents successfully self-healed earlier faults even under uncoordinated execution, achieving up to $2.4\times$ latency reduction and $\sim26\%$ cost savings. Upon closer inspection, the detailed instructions for Task 1 already contain all the load-bearing data dependencies. In Tasks~2 and~3, the limited problem diagnosis caused all planners to generate defective plans regardless of the topology. As with Task~1, parallelism consistently reduces execution time.

%Across all of our topology, planning, and agent model combinations, Task 1 proves to be the easiest, fastest to run, and also the cheapest. Thanks to the very detailed problem and solution descriptions, all agents yield a correct fix. Unsurprisingly, the wall clock time also decreases with higher parallelism. Although the overparallel plans induce some premature consumptions and stale reads, the agents can recover from them nonetheless. Contrary to expectations, the premature consumptions and stale reads do not impede the monetary cost.

%Task 2 represents the opposite extreme. Without a structured analysis of the code, no model can craft a fruitful plan proposal, regardless of the topology. None of the tests for Task 2 ever pass. In between lies Task 3, which, depending on the model and topology selection, can pass some of the tests. Although the parallel and overparallel topologies mostly reduce time, cost, and tokens spent, this only results in increased node errors, premature consumptions, and stale reads. It is precisely this kind of task that could benefit most from our envisaged \orchestrator orchestrator, as the fault is triggered by agent coordination, not by a faulty plan.

\begin{figure}[t]
\centering
\includegraphics[width=\columnwidth]{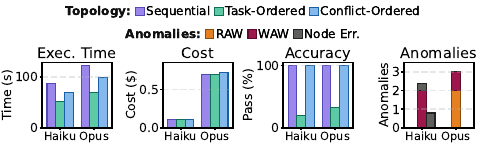}
\vspace{-5mm}
\caption{Performance of hand-written plans for Task 4.}
\label{fig:task4_evaluation_metrics}
\vspace{-3mm}
\end{figure}

\para{Repairing broken data dependencies.} In Task 4, we test how failures induced by unconstrained concurrency can be fixed (\Cref{fig:task4_evaluation_metrics}). The trivial solution to run all subtasks sequentially succeeds without errors. A task-ordered topology reduces the duration by $\sim40\%$, but only at the expense of correctness, and it does not save tokens either. Broken data dependencies induce RAW, WAW anomalies, and occasional node errors. Finally, we fed the failed plan back into the planner to identify the missed dependencies and prepare an improved conflict-ordered plan. The repaired plan was successful for both LLM agents and still reduced the task duration $1.23\times$ compared to sequential execution, at comparable token cost. These results underscore the need for an agentic orchestrator with proper data management primitives. Such primitives could even salvage verified intermediate work in the failed plan for further time and cost savings. We outline the guarantees such an orchestrator must provide in \Cref{sec:next_gen_agentic_orchestration}.

\section{The Thousand Faces of Agentic State}
\label{sec:background:data-management}
\label{sec:footprints}
The anomalies of \Cref{sec:results} arise because the state that agents read and write is scattered across components with very different properties. We distinguish three scopes of state: agent-local, orchestrator, and external (\Cref{tab:state-layers}). They differ along three axes that determine how conflicts can be detected and repaired: who can \emph{observe} the state, whether it can be \emph{reverted}, and whether it can be \emph{reproduced} after a failure. We describe each scope through Example~\ref{motiv-ex}.
%our running example (Example~\ref{motiv-ex}). %These facets of state are depicted in \Cref{tab:state-layers}.

% Resolving the workflow anomalies in Section~\ref{sec:background:control-flow} requires complementing control-flow scheduling with explicit data management primitives: how state is produced, isolated, and propagated across task boundaries. This entails explicitly tracking the storage, visibility, and updates of the propagated state. Consequently, data dependencies can be identified, and failures recovered at respective barriers. We categorize state into three operational scopes (\Cref{tab:state-layers}).

\para{Agent-local state} is private to a single agent: its context, which mutates on every iteration of the agent loop, the status of its tool calls, and, depending on the deployment, a private filesystem or key-value store. Only the agent observes this state, and its most valuable part, the context, cannot be reproduced: it is the product of prior LLM generations rather than a deterministic function of the inputs. Suppose T2 crashes while updating \texttt{contrib/auth}, after reasoning through several candidate edits but before writing any. The repository is unaffected, but nothing records \emph{what} T2 had already decided or partially produced, and its local filesystem may be left dirty. A restarted agent must either redo the reasoning or infer its predecessor's progress from the partial changes left behind.

% \begin{figure}[t]
% \centering
% \includegraphics[width=\columnwidth]{figures/state-scopes}
% \vspace{-7mm}
% \caption{State locations in a multi-agent runtime. A planner tracks workflow state; workers retain local context. Harnesses record tool execution, LLMs use derived prefix caches, and external systems hold shared effects.}
% \label{fig:state-elements}
% \Description{A planner agent contains global state, an LLM, and an orchestrator harness. Two simplified worker agents sit side by side; each contains local state, an LLM, and a harness. A shared external-systems box below lists repository files, GitHub pull requests, REST APIs, and MCP services.}
% \end{figure}

\para{Orchestrator state} comprises the cross-task runtime context maintained by the orchestrator: execution plans, task and data dependency graphs, task status, and completion signals. As the owner, the orchestrator can observe, log, and revert it at will. Some systems additionally expose a shared \emph{blackboard} that all agents read and write; since the orchestrator mediates every access, conflicts on it could, in principle, be detected.  Recent work begins to bring concurrency control to MAS workflows~\cite{lyu2026coagent,mozafari2026begin}, but to our knowledge, no orchestrator tracks conflicts over state that spans all three scopes.
% check revise
% In our example, the plan's declared \(T0 \to T2\) edge is part of the orchestration content: scheduling T2 relies on an event of T0's completion to safely proceed. Due to the not yet visible \texttt{providing\_args} attribute removal, T2's existing debug path falsely references that attribute although nothing in the ``blackboard state'' (common name for shared state across agents) records the connection between the two. The inconsistent update visibility causes the RAW violation in \ref{motiv-ex}.

\para{External state} resides outside the direct control of the orchestrator and the agents, e.g., repositories, databases, and third-party APIs. It is shared with other agents and outside consumers, and it persists after the workflow completes.  
% The repository does support staging in this case: nothing becomes externally visible until the global commit pushes the branch and opens the PR, so a conflict caught before that point can be fixed in place. Most external systems offer no such gate. 
A task that triggers a deployment through a third-party API produces an effect that is visible immediately, possibly to other consumers before any later task flags a conflict. Undoing it requires a compensating action, as in the saga pattern~\cite{garcia1987sagas}; retrying it is safe only if the operation is idempotent.
In both cases, external writes must be tracked for conflict detection. % at all.

\para{Agent state footprints.} The three scopes suggest an underlying model. Each task operates with a read set and a write set comprising items from any scope: a repository file, the agent's context itself, a blackboard entry, or a tool call result. Together, they form the task's \emph{state footprint}. Two tasks conflict if one writes an item that the other reads or writes, and a schedule is \emph{conflict-ordered} if every pair of conflicting tasks is ordered. 
% The Task-Ordered plan of \Cref{sec:results} failed this test: \texttt{providing\_args} $\in W(\text{T0}) \cap R(\text{T2})$, yet the conflict spans two scopes: the write happened in the repository, while the read happened when T2 loaded the definition into its context. No single-scope mechanism, whether git, a sandbox, or a blackboard, can observe both sides. 
Tracking an agent's footprint is hard because the read and write sets are not fixed a priori. An agent's read/write set is materialized in its context window, and new dependencies may emerge during execution. A sound approximation is to record everything an agent observes through its tool calls and task inputs. Interestingly, one can treat agent searches like predicate reads. Since agents can perform \emph{semantic recovery} rather than rolling back to the state before a failed or conflicting task, an agent may be able to self-heal from a half-written file or a conflicting state mutation. But the orchestrator needs to have a complete picture of what state has been mutated, and this can be provided by our proposed state footprints.

\begin{table}[tb]
\centering
\rowcolors{2}{white}{gray!30}
\caption{State scopes in MAS, and their properties.}
\vspace{-2mm}
\label{tab:state-layers}
\footnotesize
\setlength{\tabcolsep}{2pt}
\begin{tabularx}{\columnwidth}{@{}>{\raggedright\arraybackslash}p{42pt} >{\raggedright\arraybackslash}X >{\raggedright\arraybackslash}p{44pt} >{\raggedright\arraybackslash}p{42pt} >{\raggedright\arraybackslash}p{44pt}@{}}
\specialrule{.1em}{.05em}{.05em}
\textbf{Scope} & \textbf{State} & \textbf{Observable by} & \textbf{Reversible} & \textbf{Repro\-ducible} \\
\specialrule{.1em}{.05em}{.05em}

Agent-local &
context, tool calls, unstaged files &
agent only & via sandbox/ snapshot & artifacts, not context \\

Orchestration &
task graph, status, idempotency IDs &
orchestrator & yes (owned) & yes (logged) \\

External &
repos, REST APIs, MCP, databases &
everyone & if staged; else compensate & only if idempotent \\

\specialrule{.1em}{.05em}{.05em}
\end{tabularx}
\vspace{-5mm}
\end{table}

\section{Vision and Research Challenges}\label{sec:next_gen_agentic_orchestration}

Making MAS reliable requires rethinking how transactional database guarantees meet the unique challenges of MAS. Existing work addresses individual pieces: isolation~\cite{mozafari2026begin,lyu2026coagent}, fault tolerance~\cite{asthana2026structured}, and replanning~\cite{kaoudi2026rethinking}, but none treat state management holistically across all three considered scopes. We argue that next-generation orchestrators should generalize: the state footprint of each task is declared or inferred (\S\ref{sec:footprints}), and the orchestrator records the provenance linking each write to the tasks that consumed it. This separates nondeterministic LLM behavior from deterministic guarantees: the plan and task actions come from LLMs, which sample outputs and may be wrong, but the guarantees come from a deterministic orchestrator that records footprints and enforces ordering regardless of LLM behavior. We outline this vision via well-established ACID concepts and identify open research questions.

\subsection{Atomicity and Reversibility}\label{sec:atomicity}

%Consistency of Global state. Orchestrator that needs all agents to make their changes to code at once (coordinated commit) instead of individually (local commits). Ability to reverse/rollback. But we already know about those. 

%How about cases where consistency goes beyond read-write sets? e.g., changes in two files can conflict but a merger agent can fix those conflicts. The same in a database. 

%Parallel snapshot Isolation. (The parallel agents read and operate on the same snapshot; they commit their updates, merged together.)

A database transaction's writes target a single storage engine that can undo them. In contrast, an agent task may edit files in a local checkout, insert into a database, update configuration files, and call a web service, each with its own notion of versions, if any. Reverting any of these systems in isolation is well understood, e.g., git, filesystem sandboxes, ZFS snapshots, etc. But reverting a \emph{task} requires these heterogeneous systems to roll back together, to states corresponding to the \textit{same logical point}. Coordinated rollback has been studied for relational databases, sagas \cite{garcia1987sagas}, durable workflows \cite{aws-lambda-durable,temporal}, and cloud applications~\cite{psarakis2025styx}, but not across a mix of these systems touched by a single agent. Parallelism compounds the problem: when concurrent agents write to shared (external) state, aborting one agent must not undo the effects of another.

\para{Our vision.} Every task execution should carry a transaction identifier that propagates to every system it touches: the harness, the orchestrator, and each tool. Tool calls are recorded against this identifier together with the task's footprint (\S\ref{sec:footprints}), so that the orchestrator can abort a task by reverting every write: through a snapshot where the system supports versions, and through a compensation action where it does not. Because agents spawn sub-agents, identifiers must nest, so that a sub-task can be aborted without aborting its parent. Footprints also scope aborts under parallelism: only the aborted task's writes, and the tasks that read them, are undone, while concurrent tasks with disjoint footprints proceed.

\para{Open research questions.} (i)~How can filesystems, databases, version control, and web services agree on versioning conventions so that one transaction identifier maps to a consistent snapshot in each? (ii)~What is the right abort granularity for long-running agent tasks: the whole task, individual tool calls, or reasoning steps in between? (iii)~When should an orchestrator roll back, and when should it let an agent repair its partial effects via semantic recovery?

\subsection{Consistency and Isolation}

%\subsection{Global Agentic State Consistency.} \asterios{internal consistency and external consistency. When the orchestrator has a finished orchestration, the external systems should contain the committed state} The global agentic state refers to a global snapshot of all three types of states that we described in this section, at a given point in time. The question that arises here is: when can we consider the global state of an agentic orchestration \textit{consistent}?

%\asterios{distingsuish between single-agent local state, external state consistency, and global state consistency. Syntactioc and semantic. Post conditions, preconditions. Linearrizability is syntactic and strong. EAch agentic execution of a single agent is a subtransaction. What is a correct execution anyway? Isolation can ; Notified serializability check.}

Agents' concurrent access to shared state may cause their actions to interfere, giving rise to concurrency anomalies such as lost updates, inconsistent reads, or writes that lead to an incorrect state. The ACID notion of consistency~\cite{gray1992transaction} requires that a transaction never leaves the database in violation of its integrity constraints; a seemingly simple invariant, but very hard to get right in databases. The agentic analogue is that an orchestration \textit{must preserve the task's invariants}. Those invariants, however, vary per task. In a coding task, a trivial invariant could be that the repository must compile and pass its tests. Orchestrator state poses little difficulty here: the orchestrator owns plans, dependency graphs, and task status, and can check their invariants directly. External state is different; its invariants are implicit and checked late (e.g., at \texttt{git push}). In Example~\ref{motiv-ex}, T0's removal of \texttt{providing\_args} broke the invariant that every referenced attribute exists, yet the only check was a test suite at commit. That violation was caught because it failed loudly. However, the T1 and T2 fixture collision passed silently and could contaminate the remote repository. While the orchestrator state remained consistent throughout, the repository did not.

%However, the existing models are not directly applicable to MASs, because agentic workflows break three assumptions they rely on. (i) First, MASs have agent dependencies without a fixed vocabulary of shared objects \burcu{e.g."read set" is its broad context window. }. Unlike databases, which detect conflicts over a predeclared data model, such as rows, keys, or predicates over a known schema, an agent's dependencies are not drawn from such a set. A task may depend on a file, a function's behavior, or an API contract, which may emerge only during execution. (ii) Second, a MAS consists of nested subcomputations with their own effects, due to agents spawning other agents to perform subtasks. (iii) Third, MAS tasks are typically long-running tasks without a single commit point. Unlike database transactions, which are short and end at a well-defined commit, agentic tasks may run for minutes or hours and produce useful intermediate results along the way, which complicate the definition and application of efficient consistency and isolation models.

%The dependency graph recorded that T2 runs after T0, but not that T2's code depends on the argument T0 removed.

Isolation governs when a transaction's effects become visible to others. For external state, the answer differs by resource: a local checkout can hold writes from the remote repository but not from co-located tasks, whereas a third-party service exposes an effect immediately. Agentic isolation (in the database sense rather than sandboxing for security) also extends beyond data, e.g., acquiring a lock before triggering a deployment. The core difficulty, however, is that a task's state footprint is unknown a priori: it spans scopes and emerges only during execution. Consider a \emph{write skew}: a refactoring agent sees no references to \texttt{providing\_args} and removes its definition, while a concurrent debugging agent sees the definition and adds a reference. They edit different files and merge without a textual conflict, yet violate the invariant that every reference resolves.
%
%The example illustrates isolation holding at one external state but absent at another. The global commit node served as a commit point for the remote repository. None of the changes were pushed until the test suite passed, so remote observers never saw the inconsistent state. The tasks sharing the local filesystem had no such boundary and saw each other’s writes immediately. Thus, isolation at the remote boundary did not prevent the consistency violation, which arose in the local filesystem. For a third-party service, even the remote boundary may be absent.
%An isolation model for multi-agent systems must therefore define, for each class of external resource, when an agent's effects become observable, both to tasks sharing local state and to outside observers, and what other agents may assume about effects not yet observable.
%
A strong guarantee is serializability~\cite{papadimitriou1979serializability}, where tasks appear to execute atomically. 
However, this does not fit MASs, involving long-running, nested computations. 
%This calls for isolation models tailored to MAS, with visibility and data-dependency semantics beyond traditional models.

\para{Our vision.} Orchestrators should take responsibility for concurrency control~\cite{bernstein1987concurrency}: agents declare, or the orchestrator infers, tasks' state footprints, and the orchestrator detects schedules that are not conflict-ordered before their effects are published. Isolation should then be defined per resource class, stating when a task's effects become observable to tasks sharing local state and to outside observers. A detected conflict must invalidate both the conflicting writes and every decision derived from them\footnote{While LangGraph provides programmers with the means to handle state conflicts~\cite{langgraph2024}, it does not track the full state footprint proposed in this paper.}. Rolling back is not the only remedy: with \emph{invariant-guided semantic recovery}, agents revise invalidated decisions and repair conflicting effects, with application invariants serving as the acceptance criterion for the resulting state~\cite{lahiri2026intentformalizationgrandchallenge}. 
%removed as already cited in the beginning of the section
%Notified serializability~\cite{lyu2026coagent} and semantic isolation~\cite{mozafari2026begin} are first steps in this direction. 
We envision novel hierarchies of agent-centric consistency~\cite{DBLP:journals/csur/ViottiV16} and isolation levels~\cite{adya1999weak} tailored to MAS.

\para{Open research questions.} (i)~How can invariants over internal and external state be specified, or inferred from user intent? (ii)~At what granularity should observed reads be tracked (files, functions, or API contracts) to avoid false conflicts that needlessly limit parallelism? (iii)~What agent-centric consistency and isolation levels can we devise to navigate performance-accuracy trade-offs?

\subsection{Durability and Fault Tolerance}

In MASs, ``committed'' (i.e., durable) requires a commit that spans many systems with different guarantees: a git push, a database write, a third-party API call. The orchestrator can only acknowledge a task as complete once each system acknowledges its part. Acknowledgments must be surfaced hierarchically, from external systems to tools, through the agent harness and all the way to the orchestrator. Upon failure, the standard remedy for recovery, replaying a failed task, assumes deterministic and idempotent execution. LLM agents offer neither: re-running a task samples a different output, and its tool calls may repeat external side effects. What can be recovered, and how,  depends on the state scope (\Cref{tab:state-layers}): versioned external state can be restored from a snapshot~\cite{bernstein1983multiversion}, unversioned effects require compensation~\cite{garcia1987sagas}, derived artifacts can be recomputed, and agent context can only be regenerated by re-prompting.

\para{Our vision.} The provenance recorded alongside footprints should become part of a recovery log. On failure, the orchestrator discards only the state, causally downstream of the failed write, restores it per scope, and maintains a ledger of compensations for effects it cannot roll back. Tool calls carry idempotency keys derived from the task's transaction identifier (\S\ref{sec:atomicity}), so that a replay does not repeat side effects that have already taken place. To make replay deterministic, the orchestrator should further log every LLM output and tool response of a task, as durable execution engines log activity results in their event history~\cite{temporal}. A replayed task consumes the logged results up to the point of failure instead of resampling them. External systems, in turn, must acknowledge when an effect has become durable and visible, not merely that a request was accepted. Instead of replaying a failed agent from scratch, the orchestrator can hand the recorded footprint and partial effects to a fresh agent, in order to attempt semantic recovery. Importantly, under parallelism, tasks whose footprints are not dependent on the failed task's footprint can continue unaffected.

\para{Open research questions.} (i)~What constitutes an agent's footprint checkpoint: can we checkpoint its context at an intermediate reasoning step, so that recovery need not restart the task? (ii)~How can we extend existing recovery mechanisms to deal with heavy non-determinism? (iii)~How can durability acknowledgments from heterogeneous systems compose into a single, trustworthy commit? 

\subsection{Reuse and Dynamic Replanning}

Agents run loops: they try out branches of computation, evaluate their own progress, and revise their plans as they discover new information. Traditional systems would restart a failed task from scratch, discarding many iterations of costly reasoning. Unless the verified work of a failed run can be salvaged, a rerun re-executes every task, including those whose results were unaffected by missing dependencies. Reuse, however, requires knowing which intermediate results remain valid. A result that was checkpointed before an agent discovered a new data source, for instance, may no longer be valid, while a result whose inputs have not changed is safe to keep.

\para{Our vision.} Footprints and provenance turn reuse into a validity check. A task's result remains valid if no item in its read set changed since the result was produced, much as incremental view maintenance decides which views to refresh. Replanning then becomes a ``diff'' between the old and new plan rather than a restart. Within long-running loops, nested checkpoints in the style of Naiad's timestamps~\cite{murray2013naiad} allow an agent to resume from its last valid iteration instead of from the start. Under parallelism, a newly added dependency serializes only the conflicting tasks, so the rest of the plan keeps running concurrently.

\para{Open research questions.} (i)~When is a checkpointed agent result reusable, given that a fresh run would sample different inputs and produce a different chain of thought? (ii)~Can we build cost models to predict the trade-off between checkpointing cost and task replay? (iii)~How should checkpoints be taken in nested agent loops so that recovery resumes mid-loop without persisting every iteration?

% \subsection{One Substrate for All Guarantees}

% \begin{table}[tb]
% \centering
% \rowcolors{2}{white}{gray!30}
% \caption{Mechanisms that can enforce each guarantee, per state scope. All rely on task footprints and provenance.}
% \vspace{-3mm}
% \label{tab:guarantee-matrix}
% \footnotesize
% \setlength{\tabcolsep}{2pt}
% \begin{tabularx}{\columnwidth}{>{\raggedright\arraybackslash}p{38pt} >{\raggedright\arraybackslash}p{52pt} >{\raggedright\arraybackslash}p{46pt} >{\raggedright\arraybackslash}X}
% \specialrule{.1em}{.05em}{.05em}
% \textbf{Guarantee} & \textbf{Agent-local} & \textbf{Orchestrator} & \textbf{External} \\
% \specialrule{.1em}{.05em}{.05em}
% Atomicity & sandbox / snapshot revert & log-based undo & tid revert; try-confirm-cancel; compensation \\
% Consistency \& isolation & tool reads; predicate reads & conflict-ordered scheduling & conditional writes; branches with conflict-surfacing merge \\
% Durability \& recovery & recompute; re-prompt context & provenance log; scoped discard & hierarchical acks; idempotency keys; compensation ledger \\
% Reuse & loop checkpoints & validity on $R(T)$; plan diff & versioned reads \\
% \specialrule{.1em}{.05em}{.05em}
% \end{tabularx}
% \vspace{-5mm}
% \end{table}

\subsection{Agent-Ready External Systems}\label{sec:agent-ready}

The mechanisms above assume that external systems cooperate, yet most were built for human-paced clients whose requests are infrequent, deliberate, and mostly correct. Agents invert this assumption: they explore through speculation~\cite{liu2026overlords}, issue many requests in parallel, and many of their effects will later be aborted, retried, or compensated. An agent that mistakenly provisions virtual machines or triggers a payment leaves the orchestrator with no means of undoing the effect beyond an idempotency id and a compensation. We therefore argue that external stateful systems must be redesigned for abort-heavy, stateful workloads.

\para{Our vision.} We envision five primitives. First, external systems accept a form of transaction identifier of a task, propagate it to downstream calls, and can revert all associated effects. Second, writes are either (i)~tentative by default and take effect only upon confirmation, following the try-confirm-cancel (TCC) pattern~\cite{helland2007life,pardon2014tcc} or (ii)~compensable~\cite{garcia1987sagas}. Third, writes can be conditioned on the state that their task reads, similarly to predicated single-shot transactions~\cite{idziorek2023dynamodb}. Fourth, every operation declares whether it is read-only, idempotent, or reversible, together with its inverse. Tool annotations of MCP~\cite{mcp2025spec} are already an early step in this direction. Fifth, and most importantly, external systems offer cheap state versioning/branching, so that parallel agents can each speculate on a private copy~\cite{liu2026overlords} and merge through a step that surfaces conflicts.

\para{Open research questions.} (i)~Can we create protocols that carry transaction identifiers, read preconditions, visibility contracts, and share read/write sets across arbitrary agents and services? (ii)~Agentic transactions run for minutes or hours; how should systems bound this retention, e.g., by expiring tentative writes into automatic cancellations, without aborting legitimate long-running work? (iii)~Are novel compensation ``contracts'', apart from SAGAs and TCC, needed in order to support emerging agentic workloads?

\section{The Road Ahead}
\label{sec:looking_ahead}

The four guarantees we used here are not orthogonal: they rest on the same substrate, namely state footprints and the provenance that links each write to the tasks that read it. Atomicity uses write sets to determine what to undo. Isolation uses read and write sets to detect conflicts. Recovery uses provenance to discard only causally dependent downstream state. Reuse uses read sets to decide which results remain valid. What differs across scopes is only the guarantee enforcement mechanism, and in external state that mechanism must be exposed by the systems themselves. An orchestrator that records footprints can provide all four guarantees, and relax each independently where parallelism matters more than strictness.

We focused on coding agents as the most mature agentic use case, and because git provides a natural transactional substrate: staging, atomic commits, and cheap rollback. Most domains lack such a substrate. Supply-chain operations, scheduling, and incident response operate without version control; effects are immediately visible and often irreversible. Human-in-the-loop checkpoints offer a safeguard today, but time-critical tasks demand autonomous execution. These domains need guarantees even more than coding does, and they depend most on external systems becoming agent-ready.

One may argue that better models can make these concerns obsolete. We argue for the opposite: stronger models will be trusted with longer tasks, more parallel agents, and more consequential side effects, which enlarges exactly the state that must be coordinated. Coordination, isolation, and recovery are properties of the system, not of the model. The data management community has spent decades of researching methods and abstractions. Agentic orchestration is where they are needed next.

%\input{sections/07_tx_agents}
%\input{sections/related_work}
%\input{sections/road_ahead}

%%
%% The acknowledgments section is defined using the "acks" environment
%% (and NOT an unnumbered section). This ensures the proper
%% identification of the section in the article metadata, and the
%% consistent spelling of the heading.
%\begin{acks}
%This publication is part of project number 19708 of the Vidi research program, which is partly financed by the Dutch Research Council (NWO).
%\end{acks}

%%
%% The next two lines define the bibliography style to be used, and
%% the bibliography file.
\bibliographystyle{ACM-Reference-Format}
\bibliography{references.bib}

%%
%% If your work has an appendix, this is the place to put it.
\appendix

\end{document}